\documentclass{INTERSPEECH2023}
\usepackage{amsmath,amsfonts}
\usepackage{amsmath}
\usepackage{algorithmic}
\usepackage{algorithm}
\usepackage{array}
\usepackage[caption=false,font=normalsize,labelfont=sf,textfont=sf]{subfig}
\usepackage{textcomp}
\usepackage{stfloats}
\usepackage{url}
\usepackage{verbatim}
\usepackage{graphicx, multirow}
\usepackage{cite}
\usepackage{pifont}

\usepackage{nicematrix}
\usepackage{makecell}
\newcolumntype{C}[1]{>{\centering\arraybackslash}p{#1}}

 \interspeechcameraready

\title{Self-supervised learning with diffusion-based multichannel speech enhancement for speaker verification under noisy conditions}
\name{Sandipana Dowerah$^*$, Ajinkya Kulkarni$^{\dagger}$, Romain Serizel$^*$, Denis Jouvet$^*$}
\address{
  $^*$Université de Lorraine, CNRS, Inria, Loria, F-54000, Nancy, France, $^{\dagger}$MBZUAI University, UAE}
\email{sandipana.dowerah@loria.fr, ajinkya.kulkarni@mbzuai.ac.ae, romain.serizel@loria.fr, denis.jouvet@inria.fr}

\begin{document}

\maketitle
 
\begin{abstract}

The paper introduces Diff-Filter, a multichannel speech enhancement approach based on the diffusion probabilistic model, for improving speaker verification performance under noisy and reverberant conditions. It also presents a new two-step training procedure that takes the benefit of self-supervised learning. In the first stage, the Diff-Filter is trained by conducting time-domain speech filtering using a scoring-based diffusion model. In the second stage, the Diff-Filter is jointly optimized with a pre-trained ECAPA-TDNN speaker verification model under a self-supervised learning framework. We present a novel loss based on equal error rate. This loss is used to conduct self-supervised learning on a dataset that is not labelled in terms of speakers. The proposed approach is evaluated on MultiSV, a multichannel speaker verification dataset, and shows significant improvements in performance under noisy multichannel conditions.
\end{abstract}

\section{Introduction}

Speaker verification (SV) aims to confirm the identity of a person based on his/her voice characteristics. SV has achieved significant performance gain in controlled or close-talk scenarios. However, it suffers from unsatisfactory performance in multichannel far-field scenarios. This is due to complex environmental settings as speech signals propagating in the long-range are subject to fading, absorption, room reverberation and complex environmental noises, which change the pressure level at different frequencies and degrade the signal quality. Speech enhancement (SE) can be used as a pre-processing to SV in noisy reverberant scenarios. Speech enhancement aims to enhance the quality and intelligibility of speech signals that are corrupted by noise and/or reverberation by estimating the original clean speech signal using various signal processing techniques. Multichannel speech enhancement aims to enhance distorted speech using multiple microphones and improve performance by taking advantage of the additional spatial information provided by these microphones compared to single-channel.

%The problem is likely to arise in the SV part, where we need the speaker Id annotations. Using speech enhancement as the front end to a far-field SV could solve this. When fine-tuning the ground truth of speech enhancement is not needed, but multichannel data with speaker annotation is needed, which could be a problem. 

Generative models aim to learn the fundamental characteristics of speech, such as its spectral and temporal structure and can use this prior knowledge to identify clean speech from noisy or reverberant input signals that fall outside the learned distribution. \cite{Pascual2017SEGANSE, Michelsanti2017ConditionalGA} used the raw waveform, or magnitude spectrum, as input for generative model-based speech enhancement. Generative adversarial networks (GAN) \cite{Soni2018TimeFrequencyMS, Fu2019MetricGANGA}, variational autoencoders (VAE) \cite{Bando2017StatisticalSE, Leglaive2018AVM, Bando2020AdaptiveNS}, and flow-based models \cite{Nugraha2020AFD} have been used to estimate the distribution of clean speech signals. Recently, diffusion-based models have also been studied for speech enhancement \cite{diffuSE, ConditionalDP, Kong2020DiffWaveAV}. All these approaches share the concept of gradually converting input data into noise and training a neural network to invert this process for various noise scales based on the Markov chain. 

DiffuSE \cite{diffuSE} was proposed to recover the clean speech signal from the noisy signal based on Markov chains; it provides a framework for denoising diffusion probabilistic models. Lu et al. formulated the CDiffSE model using a generalized conditional diffusion probabilistic model that incorporates the observed noisy data into the model \cite{ConditionalDP}. While CDiffSE and DiffSE employ U-net as their diffusion decoder network, our proposed work takes a different approach and uses Conv-TasNet as the diffusion decoder instead. Specifically, our method conducts speech enhancement on the time-domain representation of the signal. Zhang et al. extend the Diff-Wave vocoder \cite{Kong2020DiffWaveAV} using a convolutional conditioner for denoising, and it is trained separately using a L$1$ loss for matching latent representations \cite{Zhang2021RestoringDS}.Our proposed approach incorporates a conditioning network based on Conv-TasNet in addition to the diffusion decoder. This conditioning network provides an estimate of the clean and noisy signals, which are combined with the multichannel noisy signal and fed into the diffusion decoder. By doing so, the diffusion process is made easier as it can learn to remove the noise while taking into account the clean speech estimate provided by the conditioning network. Recently, some studies \cite{dowerah:hal-03671583, serra2022universal, welker2022speech} have explored scoring-based diffusion models with stochastic differential equations (SDE) instead of Markov chains. SDE enables the controlling of selecting the reverse diffusion steps for enhancement \cite{Song}. The aforementioned works use only single-channel and have not been studied for SV. 

% SSL then why eer loss and conditoning network importance

% An augmentation adversarial training was introduced in \cite{Kang2022AugmentationAT} to extract the channel-invariant speaker representations. The distillation with no labels (DINO) was used to train SV with dynamic loss-gate and label correction strategy \cite{Han2022SelfSupervisedSV}.

Self-supervised learning is a powerful machine learning technique that enables models to learn from unlabeled data by leveraging the inherent structure or patterns in the data itself without the need for explicit supervision from labelled data. In the context of speaker verification tasks, few approaches have conducted contrastive learning for self-supervised learning \cite{Kang2022AugmentationAT,Han2022SelfSupervisedSV,Xia2020SelfSupervisedTS,Nagrani2020DisentangledSE}. The loss function design for SV mainly focuses on speaker classification loss function, and verification loss \cite{Mingote2021LogLikelihoodRatioCF}. Furthermore, the contrastive learning framework enables the online creation of verification labels. In order to exploit the multichannel speech data without explicit speaker labels, we propose to use the equal error rate (EER) evaluation metric as a loss function to optimize the speaker embedding representation on the verification task.

In this paper, we present a diffusion probabilistic model (DPM)-based two-stage multichannel speech enhancement approach as a pre-processing to SV. We named our approach Diff-Filter as it mimics the behaviour of Rank-1 multichannel Wiener filter (MWF). In the first stage, we train the Diff-Filter by conducting time-domain speech filtering using a scoring-based diffusion model. In the second stage of training, we jointly optimize the Diff-Filter with a pre-trained ECAPA-TDNN SV model under a self-supervised learning framework. We evaluate our results on MultiSV, a multichannel SV dataset, and show that our proposed approach significantly improves SV performance under multichannel noisy conditions.

%Dowerah et al. illustrated that Rank-1 MWF improves SV tasks compared to using other filtering techniques, such as GEV, or beamforming techniques, such as MVDR \cite{aspai}. 

%The Diff-Filter leverages a Conv-TasNet \cite{Luo2018ConvTasNetSI} decoder network to generate a Rank-1 multichannel Wiener-filtered (MWF) \cite{r1mwf} clean speech signal for a given input as a noisy multichannel signal. We incorporate a conditioning network based on Conv-TasNet to provide estimates of clean speech and noise as inputs to the diffusion-based decoder network, mimicking the behaviour of Rank 1-MWF. 

% problem description

% se and diffusion approaches and their limits

% work done on se for sv

% work done with ssl but not with multichannel se sv

% highlights of proposed work

\section{Proposed Approach}

% general description of three-stage approach, 
% 1. training SE
% 2. training sv
% 3. training jointly with SSL using eer loss
% 4. additional train set using libriSpeech and musan noise augmentation

In this section, we present the proposed approach for developing a robust multichannel SV system in a noisy environment. In the first phase, we trained the ECAPA-TDNN \cite{Desplanques2020ECAPATDNNEC} based SV system and a multichannel speech enhancement system separately. We used this pre-training of speech enhancement and SV for training the jointly optimized system using self-supervised learning. We jointly optimized Diff-Filter and ECAPA-TDNN with the proposed EER loss as a verification loss to optimize the binary classification with speaker embedding representations. Diff-Filter is a scoring-based diffusion probabilistic model where Conv-TasNet architecture is utilized for conducting the diffusion process. Diff-Filter is trained to provide Rank-1 MWF clean speech signal for a given multichannel noisy input signal. We used a conditioning network to provide the estimates of clean and noise signals as additional input to the diffusion decoder, thus conditioning the sampling process from terminal distribution aware of noise to be removed from the noisy multichannel signal. 

\subsection{Diff-Filter}

\begin{figure}[!t]
    \centering
    \includegraphics[width=\columnwidth]{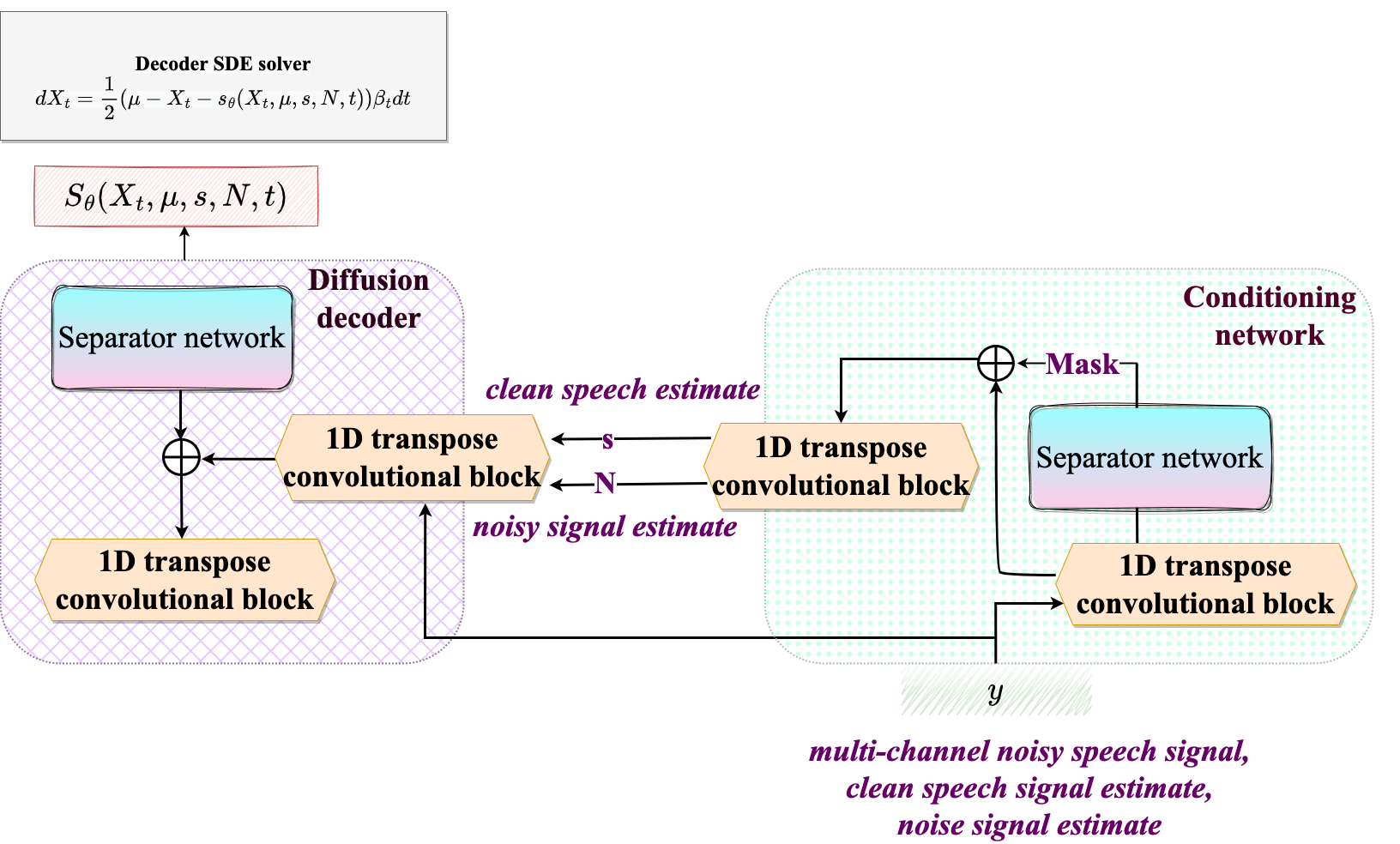}
    \caption{Architecture of Diff-Filter consists of a diffusion decoder and a conditioning network. The clean speech signal and noise signal estimated from the conditioning network are channel-wise concentrated with noisy multichannel signal along with Rank-1 MWF clean speech signal in the diffusion decoder.}
    \label{fig:diff-fil}
    \vspace{-2em}
   
\end{figure}

This section presents a novel way to train a multichannel speech enhancement system as a DPM-based filtering method named Diff-Filter. We termed the proposed system Diff-Filter, as it replicates the functionality of the Rank-1 MWF filter to provide a clean speech signal. The proposed Diff-Filter comprises a diffusion-based decoder network and a conditioning network, as shown in Figure \ref{fig:diff-fil}. We used Conv-TasNet \cite{Luo2018ConvTasNetSI} as an external conditioning network. The conditioning network is used to compute the estimates of the clean speech signal, $s$, and noise in time-domain representation, $N$. We provide conditioning network output estimates along with the multichannel noisy speech signal as input to the Diff-Filter system. In the forward and reverse diffusion process, terminal noise distribution is defined as $\mathcal{N}(\mu, I)$, where the mean is $\mu$, and $I$ is unit variance. We parameterized the mean $\mu$ of terminal noise distribution of the diffusion process using noisy multichannel input, $y$. 

%We use SI-SDR loss to estimate the clean speech and noise to be given as additional input to the diffusion decoder.

Similar to \cite{dowerah:hal-03671583}, we incorporated scoring-based diffusion probabilistic model, in which the diffusion decoder learns the trajectories of forward diffusion in reverse time order. In the training phase, the forward diffusion process is conducted by iteratively deconstructing Rank-1 MWF clean speech estimate signal to the terminal distribution defined by noisy multichannel signal. Furthermore, terminal distribution is also conditioned with estimates of the clean speech signal and noise signal provided by the conditioning network. The usage of clean speech estimate and noise estimate in the diffusion process assists in conducting noise-aware speech enhancement. We used stochastic differential equations (SDE) to learn the gradients of the forward diffusion process as shown in Figure 1, where $s_{\theta}(X_t,\mu,s,N,t))$ denotes the diffusion decoder network with $t$ as the diffusion time step and $\beta$ as noise scheduler. During the inference phase solving the SDE describing dynamics of the reverse diffusion with a simple first-order Euler-Maruyama scheme \cite{Kloeden1977TheNS,Song}.

%The speech signal and noise signal are channel-wise concentrated with noisy multichannel signal along with Rank-1 MWF clean speech signal to . 

We trained the Diff-Filter using a two-stage training process. First, we conditioned the diffusion encoder with a target clean speech signal, a target noise signal, and a noisy multichannel signal. The main purpose of pre-training the diffusion decoder is to ensure that diffusion model parameters converge in optimal minima direction using target clean speech and noise. In the second stage of training, we used the clean speech and noise estimated by the conditioning network.  

In the inference phase, multichannel noisy signals and estimates of speech and noise signals obtained from the conditioning network are provided to the diffusion decoder to estimate the reverse trajectories of forward diffusion. The reverse diffusion process iteratively reconstructs the Rank-1 MWF filter output of clean speech by sampling latent variables from conditional terminal distribution. 

\subsection{Self-supervised learning for multichannel SV}

We jointly optimized the Diff-Filter and ECAPA-TDNN as a multichannel SV. In joint optimization, the multichannel noisy signal is first given to Diff-Filter, which provides a single-channel Rank-1-MWF filtered clean speech signal. The error gradient passes through ECAPA-TDNN and Diff-Filter as a single unit because we back-propagate through both models that we are jointly training. To conduct the self-supervised learning with the unlabelled dataset, we created an RIR simulated dataset and applied it to clean speech from the LibriSpeech dataset \cite{Panayotov} detailed in section 3.

%We conducted $10$ reverse diffusion steps to output the filtered clean speech signal. The output is then used for extracting the Mel spectrogram feature as input to ECAPA-TDNN, which provides the speaker embedding. 

\begin{figure}[!t]
    \centering
    \includegraphics[width=\columnwidth]{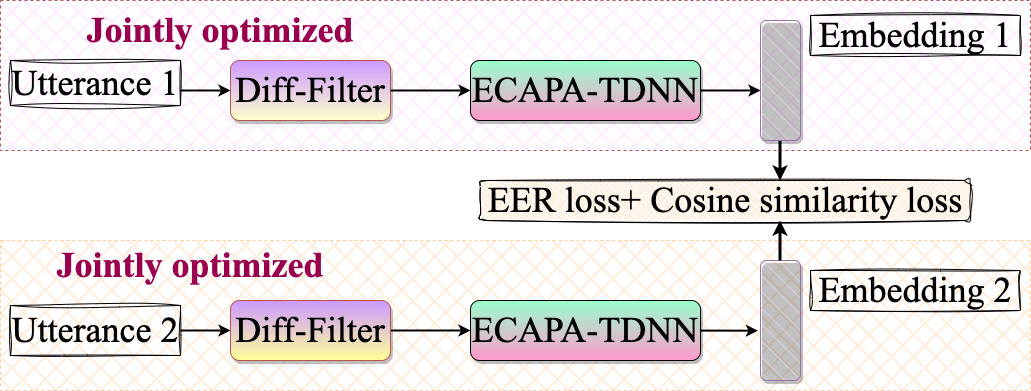}
    \caption{Framework of self-supervised learning, where $utterance_{1}$ and $utterance_{2}$ are noisy multichannel signals given to the jointly optimized network of ECAPA-TDNN to obtain the speaker embeddings.}
    \label{fig:diff-filter}
   \vspace{-1.8em}
\end{figure}

As shown in Figure \ref{fig:diff-filter}, utterances $1$ and $2$ are given to the jointly optimized network composed of Diff-Filter and ECAPA-TDNN. The utterances $1$ and $2$ are multichannel noisy signals, where utterance $2$ is either the data-augmented multichannel noisy signal from utterance $1$ or a randomly selected multichannel noisy signal from a different speaker. During the training, we used the self-supervised contrastive learning framework, where both utterances, utterances $1$ and $2$, are given to the same jointly optimized network. Verification labels are generated as $0$ or $1$ if utterance $2$ is data-augmented from utterance $2$ or not, where $1$ represents that utterances $1$ and $2$ are from the same speaker, $0$ otherwise. For data augmentation in self-supervised learning, we used speed perturbation by $0.9$, $1.1$ factor only and masking the $1$ sec part of the noisy multichannel signal. 

We propose to use an EER as a loss function to train the jointly optimized network. The EER is the location on a receiver operating characteristic curve where the false acceptance rate and false rejection rate are equal. First, we computed the cosine similarity distance between $embedding_{1}$ and $embedding_{2}$ for a given batch. Then, false acceptance rate (FAR) and false rejection rate (FRR) are estimated based on cosine scores and verification labels using torchmetrics\footnote{https://torchmetrics.readthedocs.io}. We estimated EER for the given batch size from FAR and FRR as stated in Equation 1, where $\mathcal{L}_{EER}$ ranges from value 0 to 1. 

\begin{equation}
    \centering
    \mathcal{L}_{EER} = FAR \Bigr[ argmin | FRR- FAR | \Bigr]
\end{equation}

We also estimated cosine similarity loss between embeddings: $embedding_{1}$ and $embedding_{2}$ \cite{Novoselov2018TripletLB} as shown below in Equation 2.

\begin{equation}
    \centering
    \mathcal{L}_{cosine} = 
    \begin{cases}
       1-cos(emb_{1}, emb_{2})& label=1\\    
      max(0,cos(emb_{1}, emb_{2})-M)& else \\  
    \end{cases}
\end{equation}

\noindent where $emb_1$ and $emb_2$ refers to embeddings extracted on $utt_1$ and $utt_2$, respectively, $M$ refers to the regularizer of value 0.2, and cos refers to the cosine angle between $emb_1$ and $emb_2$.

\section{Dataset Preparation}

We used various datasets at different stages while developing the proposed approach for multichannel SV in noisy conditions. We used the MultiSV dataset \cite{multisv} for training the Diff-Filter, which consists of $4$ channel speech utterances room simulated impulse response with background noises from Music, MUSAN, and freesound.org\footnote{https://freesound.org/}. The training dataset of MultiSV is simulated using the VoxCeleb2 dataset \cite{Nagrani2020VoxcelebLS}. Consistent with the Diff-Filter training data source, we utilized the VoxCeleb2 dataset with standard Kaldi-based data augmentation techniques for training ECAPA-TDNN single-channel SV. We opted for the VoxCeleb2 dataset for joint training as MultiSV is a labelled dataset, and the core of self-supervised learning is to explore the unlabelled dataset.

To jointly optimized the network, we first simulated a room impulse dataset and applied it to the clean speech from the LibriSpeech dataset without taking into account the speaker information, thus creating an unlabelled multichannel SV dataset. The pyroomacoustics toolbox\footnote{https://github.com/LCAV/pyroomacoustics} is used for room simulation with 4 channels. The room length was drawn randomly between $[3,8]$ m, the width was chosen between $[3,5]$ m, and the height was chosen between $[2,3]$ m. The absorption coefficient was drawn randomly such that the room's RT$60$ was between $[200,600]$~ms. The minimum distance between a source and the wall is $1.5$ m and $1$ m between the wall and the microphones. We generated a total of $50000$ training samples for self-supervised learning. 

To evaluate the proposed work, we used two multichannel trial protocols from the MultiSV dataset, namely MRE and MRE hard trial protocols. The evaluation set of MultiSV is retransmitted development set derived from the VOiCES dataset. In addition to MultiSV evaluation data, we also created an internal evaluation set using Fabiole corpus \cite{Ajili2016FABIOLEAS}, a French speech corpus consisting of around $6882$ audio files from $130$ native French speakers. The speech data of Fabiole has been collected from different French radio and TV shows. For creating each evaluation set, we have used $1200$ speech files from Fabiole representing $2$ hrs of evaluation material. We used the same configuration for room impulse response simulation as used for creating the training dataset for the self-supervised learning phase. We designed the evaluation set with various RIR scenarios to be used for both speech enhancement and SV. 

\section{Experimentation set-up}

%add it in this section We used the same noise scheduling strategy as used in \cite{Popov2021GradTTSAD}.

%

%After conducting quantitative experimentation, we found $10$ reverse diffusion steps as optimal. 

\subsection{Multichannel speech enhancement}

The model is trained using two loss functions, diffusion loss and scale invariant signal to distortion (SI-SDR) loss \cite{sisdr}. The diffusion loss is defined by Fisher divergence as a way to compute the scoring function, which is the gradient of change in log probability density in each diffusion step \cite{Lyu2009InterpretationAG}. The second loss function, SI-SDR loss, is applied to the output of the conditioning network to ensure that the diffusion model ingrains the intrinsic information about clean speech estimate and noise estimate in time-domain representation. In training, we provided speech segments of a fixed length of $4$ seconds of duration.

We set the initial weight of $0.001$ on SI-SDR loss. Then, we increased the initial weight by $0.0001$ after every $5$ epoch till it reached $1$. For the two-stage training approach, first, we trained the network for $100$ epochs with a learning rate of 1e-2 and reduced the learning rate over the epochs with a factor of $0.85$ after every $5$ epoch. We used Adam optimizer for two-stage training with a batch size of $2$. In the second stage of training, the system is trained with a learning rate of $1e-4$ for $500$ epochs.

We used Conv-TasNet architecture to develop both diffusion decoder and conditioning network, with modification of replacing PReLU activation function with GeLU \cite{Hendrycks2016GaussianEL}. The implementation of networks using Conv-TasNet includes $512$ filters in the convolutional block and transpose convolutional block (N), $20$ lengths of filters (L), $256$ channels in a bottleneck, and the residual paths $1\times1$ convolutional blocks. Each convolutional block's kernel size (P) is set to $3$, and the number of convolutional blocks in each repeat is $8$. Also, we adopted global layer normalization with a non-causal strategy for Diff-Filter implementation. To ensure a stable learning process, we used gradient clipping with a maximum L$2$-norm of $5$.

We conducted self-supervised training on the proposed approach in a contrastive learning framework for $50$k iterations with a batch size of $4$. In each batch of self-supervised training, we kept equal distribution of verification labels as 0 and 1. We used Adam optimizer with a learning rate of $1e-3$ with weight decay of $1e-4$ for every $1000$ iteration.

\subsection{Speaker verification}

We used ECAPA-TDNN as a single-channel SV system from \cite{Desplanques2020ECAPATDNNEC}. We used the VoxCeleb2 dev dataset for training ECAPA-TDNN. As SV systems often benefit from data augmentation, we used a combination of different data-augmentation techniques, such as Kaldi recipes of data-augmentation (using MUSAN \cite{musan} and room impulse response dataset\footnote{https://www.openslr.org/28/}) and speed perturbation by changing the tempo of speech. 

Besides squeeze and excitation block, the attention module of ECAPA-TDNN is set to $128$. The scale dimension in Res2Block is set to $8$. We extracted $256$ dimension speaker embedding from the ECAPA-TDNN network. Initially, we trained the ECAPA-TDNN network with a cyclic learning rate varying between $1e-8$ and $1e-3$ using the triangular policy with Adam optimizer. The ECAPA-TDNN network is trained with angular margin softmax with a margin of $0.3$ and softmax pre-scaling of $30$, $100$k iterations. We provided the Mel spectrogram as an input to ECAPA-TDNN. We extracted $40$-dimensional Mel spectrogram features using the torchaudio library with a window length of $400$ samples, hop size of $160$, and $512$ FFT length/ Mel spectrogram features of $40$ dimensions as input to the ECAPA-TDNN network. We used a cosine scoring system for verification purposes from extracted embedding.

%Hence, we use the pretext task learning from labelled SV dataset to facilitate the self-supervised learning for developing multichannel noise robust SV system. 

\begin{table}[!t]
\centering
\caption{Evaluation of proposed approach on MultiSV dataset for MRE and MRE hard as multichannel trial protocol, where J. op. refers to a jointly optimized system, and SSL refers to the system trained using self-supervised learning.}
\resizebox{\columnwidth}{!}{%
\begin{tabular}{|l |c| c| c| c|} 
\hline & \textbf{SE} & \textbf{SV} & \textbf{MRE} & \textbf{MRE hard} \\
 \hline 

 {} &  Mask \cite{multisv} 			& Resnet 		& 3.91 & 5.37  \\ 
 {} &  ConvTasNet \cite{multisv}		& Resnet 		& 3.71 & 4.61   \\
 {} &  Unprocessed 		& ECAPA-TDNN 	& 5.84 & 10.27   \\  
 {} &  Oracle Rank1-MWF 	& ECAPA-TDNN 	& 1.64 & 3.12 \\  
 {} &  ConvTasNet 		& ECAPA-TDNN 	& 3.73 & 4.52 \\ 
 {} &  Diff-Filter 		& ECAPA-TDNN 	& 3.57 & 4.36 \\ 
 \hline \hline
    \parbox[t]{3mm}{\multirow{2}{*}{\rotatebox[origin=c]{90}{J. op.}}} & Diff-Filter & ECAPA-TDNN  & 3.24 & 4.26 \\
   {} & Diff-Filter & ECAPA-TDNN (SSL) & \textbf{3.07} & \textbf{3.19} \\
   
\hline
\end{tabular}
}
\vspace{-0.7em}
\label{tab:stat}
\end{table}

\section{Results and Discussion}

We compared the performance of the proposed approach with Conv-TasNet as baseline multichannel speech enhancement used as a front end to the ECAPA-TDNN system. For establishing baseline Conv-TasNet, we trained under the same training data used by the Diff-Filter system. Also, we used the same network configuration for Conv-TasNet as for the conditioning network of Diff-Filter. In addition to this, we also computed performance with oracle Rank-1 MWF in order to analyze the filtering approach based on the diffusion probabilistic model. We used EER as an evaluation metric to evaluate the multichannel SV systems on MRE and MRE hard trials from the MultiSV dataset. We compute signal-to-inference ratio (SIR), signal-to-distortion ratio (SDR), and EER on a Fabiole-based multichannel evaluation set. We used MIR eval tool\footnote{https://craffel.github.io/mir\_eval/} to compute the SIR and SDR metrics. The usage of SIR and SDR metrics provides insight into the performance of the multichannel speech enhancement system as a front end to the SV system. 

In Table \ref{tab:stat}, the Diff-Filter front-end outperforms the Conv-TasNet without additional post-training using joint optimization or self-supervised learning. We observed that the proposed approach showed better results on both trials MRE and MRE hard compared to baseline results presented in \cite{multisv}, where the Resnet-based SV system was used. We obtained the best results on the proposed approach trained under a self-supervised learning framework, which shows an efficient generalization of speaker representation under noisy conditions using an unlabelled speaker dataset. In the case of the MRE hard protocol, it has performance close to the multichannel speech enhancement baseline using Oracle Rank-1-MWF. On the other hand, the performance of the proposed approach had a significant margin in performance difference with oracle Rank-1 MWF. Table \ref{tab:stat1} illustrates consistent performance improvement by the proposed approach on both trials sets on the Fabiole-based evaluation set. SDR and SIR seem to be closely co-related with EER. With a SIR of $24.37$, the proposed joint optimized approach with self-supervised learning achieves the best performance among all the speech enhancement systems. Similarly, with an SDR of $7.02$, the proposed joint optimization approach with self-supervised learning achieves the best performance among all the systems evaluated. SIR and SDR

The proposed approach shows consistent performance on both SV and multichannel speech enhancement tasks. The usage of self-supervised learning eases the network optimization for generalization from the unlabelled distribution. As one of the primary evaluation metrics for the SV task is EER, the adaptation of EER loss without speaker labels in self-supervised training elevates the intraclass speaker representation while increasing the interclass speaker representation. The usage of a conditioning network allowed the diffusion process allowed to perform a noise-aware reverse diffusion process. The usage of Conv-TasNet as a diffusion decoder enabled to perform the step-wise noise removal on time-domain signal representation, thus inherently considering the phase information.

\begin{table}[!t]

\centering
\caption{Evaluation of proposed approach on room impulse simulated data on Fabiole dataset, where we used ECAPA-TDNN as SV system and J. op. refer to the jointly optimized system, and SSL refers to the system trained using self-supervised learning.}
\begin{tabular}{|l |c| c| c| c|} 
\hline & \textbf{SE System} & \textbf{EER} & \textbf{SIR} & \textbf{SDR} \\
 \hline 

  {} & Unprocessed 	& 9.23	& 15.11	& 2.01 \\ 
  {} & Oracle Rank-1 MWF 	& 5.91	& 24.73	& 7.24 \\ 
  {} & ConvTasnet 	& 7.87	& 23.21	& 6.12 \\  
  {} & Diff-Filter 	& 7.83	& 23.78	& 6.69 \\ 
   \hline \hline
  \parbox[t]{2mm}{\multirow{2}{*}{\rotatebox[origin=c]{90}{J. op.}}} & Diff-Filter & 7.54 & 24.11	& 6.93 \\  
  {} & Diff-Filter (SSL)        & \textbf{6.27}	& \textbf{24.37}	& \textbf{7.02} \\ [1ex] 
 \hline
\end{tabular}
\vspace{-0.7em}
\label{tab:stat1}
\end{table}

\section{Conclusion}

In this work, we proposed Diff-Filter, a multichannel speech enhancement approach as a front end to SV. We improved the performance of the proposed Diff-Filter by jointly optimizing it with ECAPA-TDNN-based SV and further training under self-supervised contrastive learning. We presented EER loss in self-supervised learning to exploit the unlabelled speaker dataset. The obtained results have shown significant improvement in performance on the MultiSV dataset compared to state-of-the-art systems. In order to measure speech enhancement performance, we used SIR and SDR evaluation metrics. The results computed on the simulated evaluation set (derived from Fabiole) showed results in-line with performance on the MultiSV evaluation set. In future, we will conduct further experimentation with Diff-Filter to observe the efficiency of different tasks such as source separation, speaker diarization etc.

\bibliographystyle{IEEEtran}
\bibliography{mybib}

\end{document}